\documentclass{mem}
\usepackage{natbib}
\usepackage{txfonts}
\usepackage{balance}
\usepackage{graphicx}
\usepackage[a4paper]{hyperref}

\def\specchar#1{\uppercase{#1}} 
\def\CaII{\mbox{Ca\,\specchar{ii}}} 
\def\Caline{\CaII\ 8542~\AA} 
\def\Halpha{\mbox{H\hspace{0.1ex}$\alpha$}}
\def\rmit#1{{\it #1}}  
\def\eg{\rmit{e.g.,}}
\def\rmH{{\rm H}}
\def\Hmin{\hbox{\rmH$^{^{_{\scriptstyle -}}}$}}

\begin{document}

\title{Numerical simulations of the quiet chromosphere}

\subtitle{}

\author{Jorrit Leenaarts\inst{1}}

\offprints{J. Leenaarts}
 
\institute{Astronomical Institute - Utrecht University - PO Box
  80\,000 NL--3508 TA Utrecht, The~Netherlands
\email{j.leenaarts@uu.nl}}

\authorrunning{Leenaarts}

\titlerunning{Simulations of the quiet chromosphere}

\abstract{ Numerical simulations of the solar chromosphere have become
  increasingly realistic over the past 5 years. However, many observed
  chromospheric structures and behavior are not reproduced. Current
  models do not show fibrils in \Caline, and neither reproduce the
  \Caline\ bisector. The emergent \Halpha\ line core intensity computed
  from the models show granulation instead of chromospheric shocks or
  fibrils. I discuss these deficiencies and speculate about what
  physics should be included to alleviate these shortcomings.
  \keywords{Sun: chromosphere, Methods: numerical} } \maketitle{}

\section{Introduction}

The solar chromosphere is in many ways the most complicated and most
interesting region of the solar atmosphere.

The chromosphere is the interface layer between the convection
zone/photosphere and the corona where the value of the plasma
$\beta$-parameter changes from larger to smaller than
unity. Consequently, its overall character changes from a convective
overshoot layer into a regime where the three-dimensional (3D)
magnetic topology determines the structure. The chromosphere is
convectively stable, so the dominant input of mechanical energy comes
in the form of waves excited by granules and the buffeting of magnetic
concentrations in the photosphere. It also absorbs radiation from
both the photosphere (\eg\ \Hmin, Balmer-continuum) and the corona (He
continua). Thermal conduction in the corona has a strong influence on
the shape of the transition region, and hence of the upper
chromosphere.  The dominant chromospheric energy loss is through
radiation in strong lines.  Thus, a comprehensive model of the
chromosphere cannot be made without modeling the underlying
photosphere and upper convection zone and the overlying lower corona.

The wide range of pertinent physical processes require that a
comprehensive description of the chromosphere needs to incorporate a
careful combination of atomic and molecular physics, radiative
transfer, MHD and plasma physics. This complicated physics requires
numerical modeling.

From the above discussion I conclude that a ``minimally satisfying
self-consistent model'' of the chromosphere should
\begin{itemize}
\item be time-dependent;
\item be three-dimensional;
\item model radiative heating and cooling;
\item have a lower boundary in the convection zone, with a boundary
  condition that feeds enough energy into the simulation to maintain a
  the radiative energy loss at the solar effective temperature;
\item have an upper boundary in the corona. If the model cannot
  self-consistently maintain a corona, it must provide enough energy
  flux to keep plasma near the boundary at coronal temperatures.
\end{itemize}
Within these constraints one can add features and physics to the model
to increase its realism, while at the same time trying to keep the
number of free parameters to a minimum.

The above list does not imply that models without those properties are
somehow ``bad models''. To the contrary, simpler models are often
instrumental in identifying and characterizing relevant physical
processes. However, a thorough understanding of the chromosphere
ultimately requires a self-consistent model that reproduces all
observed chromospheric behavior.

\section{Current status}

The current state-of-the-art in self-consistent chromospheric
modeling is the modeling performed with the
radiation-magneto-hydrodynamics (RMHD) Oslo Stagger Code
\citep[OSC][]{2004IAUS..223..385H} 
and the MPI-parallelized Bifrost code currently under development
\citep{gudiksen++2010}. 

These codes are based on the methods pioneered by 
\citet{1982A&A...107....1N}
in 3D convection simulations. They use the single-fluid MHD
approach to model the solar gas and magnetic field. Radiative transfer
in the photosphere and lower chromosphere is treated using multi-group
opacity binning with scattering
\citep{2000ApJ...536..465S}
with typically 4 radiation bins. This multi-group scheme is not well
suited to model the few strong chromospheric lines that cause the
dominant radiative energy losses in the upper chromosphere. Therefore
both codes employ parameterized radiative cooling in those lines in the
upper chromosphere based on the detailed radiative transfer
in a 1D dynamical chromospheric model
\citep[computed with the RADYN code, see \eg\ ][]{2002ApJ...572..626C}.
Radiative losses in the corona are computed using an optically thin
radiative loss function. Heat conduction parallel to the magnetic
field is included. Two-dimensional simulations cannot maintain a
corona by themselves and require additional heating at the upper
boundary. Three-dimensional models however self-consistently maintain
a corona by Joule heating.

Similar codes exist, such as Muram 
\citep{2005A&A...429..335V}, 
CO$^5$BOLD
\citep{2002AN....323..213F,2004A&A...414.1121W}
and RADMHD
\citep{2008ASPC..383..327A}. 
However these codes so far do not support a corona (CO$^5$BOLD, Muram)
or lack detailed treatment of the radiation field (RADMHD).

OSC has been used to study dynamic fibrils (DFs) by
\citet{2006ApJ...647L..73H}
and
\citet{2007ApJ...655..624D}. 
They conclude that DFs are driven by magneto-acoustic shocks based on
the similarity in dynamic behavior of wave-guided shocks in a 2D
simulation and observations of DFs in the core of H$\alpha$. Their
work has been extended to 3D in simulations by
\citet{2009ApJ...701.1569M}.

3D simulations of flux emergence from the convection zone through the
chromosphere into the corona were performed by
\citet{2008ApJ...679..871M,2009ApJ...702..129M}
A similar simulation showed the presence of Alfv\'enic waves with
properties very much like what has been observed with the Hinode
spacecraft
\citep{2007Sci...318.1574D}.

For a good overview of 3D solar atmosphere modeling and
post-processing in the form of 3D radiative transfer I refer the
reader to
\citet{2008PhST..133a4012C,2009MmSAI..80..606C}
and references therein.

Despite the impressive progress that has been made in the last 5
years, many, often surprisingly basic properties of the chromosphere
are not yet reproduced in the models. In the next section I discuss
some of those observational facts and speculate about what is missing
in the models.

\section{Open questions and challenges}

\subsection{Lack of extended magnetic structure in \Caline}

The first thing one notices in \Caline\ line-core images are the
elongated structures delineating magnetic field lines around network
regions. The NLTE \Caline\ computation from a snapshot of a 3D
simulation of the chromosphere with OSC of
\citet{2009ApJ...694L.128L}
does, however, not show any such structures, despite the presence of
a relatively strong magnetic field. Instead it shows the signature of
oscillations and shocks, both in relatively field-free regions as in
regions with stronger magnetic fields. Similar structures are observed
in \Caline\ quiet regions away from the network. The lack of fibrils
in the simulation is in my opinion the most pressing challenge models
of the chromosphere currently face.

There can be many reasons for this lack. It might be that such fibrils
require large-scale magnetic field configurations that do not fit in
the typical size of the computational domain of $16 \times 16 \times
16$~Mm$^3$. It might be that the simulation cannot form fibrils due to
a lack of grid resolution. Alternatively, fibril-like structures might
exist in the models, but the lack of resolution inhibits mass loading
of the fibrils, possibly due to too large numerical viscosity (see
also Sec.~\ref{leenaarts_sec:brobis}). This lack of mass then translates into
optically thin fibrils. Another possibility is that \Caline\ opacity
in the fibrils is caused by physical processes currently not modeled
in the simulations, for example non-equilibrium ionization or plasma
effects that cannot be modeled within the MHD assumption.

\subsection{Line broadening and bisector} \label{leenaarts_sec:brobis}

The spatially averaged \Caline\ line profile computed in
3D NLTE from the same OSC snapshot shows a too narrow and too deep line
core compared to observations
\citep[see Figs.~2 and~3 of][]{2009ApJ...694L.128L}.
These calculations did not employ any non-physical microturbulence,
and had a rather large grid cell size of
$64\times64\times32$~km$^3$. The too narrow core was attributed to
this lack of spatial resolution. Indeed, a newer simulation by
Carlsson (private communication) with a cell size of
$32\times32\times16$~km$^3$ shows a shallower and wider average
\Caline\ core, but not yet as wide and shallow as observed. The
increase in grid resolution causes an increase in amplitude of the
velocity variations in the mid and upper chromosphere, and hence a
widening of the average profile.

This result suggests that with even higher resolution the width of the
line core might at least be qualitatively explained. However, the
observed profile is at least partially formed in fibrils and as long
as the models do not reproduce these the core width and depth remains
not completely understood.

The low resolution simulation also did not reproduce the observed
inverse C-shape of the \Caline\ line bisector. Similar results were
obtained earlier by
\citet{2006ApJ...639..516U}
who computed the NLTE line profiles from both a 3D simulation
that did not extend into the corona and a dynamic 1D model. It would
be interesting to see whether the bisector shape can be reproduced by
high-resolution simulations, or that it is determined by phenomena
currently not present in the simulations.

\subsection{Lack of \Halpha\ opacity}

 \begin{figure*}
\centering
\includegraphics[width=8.8cm]{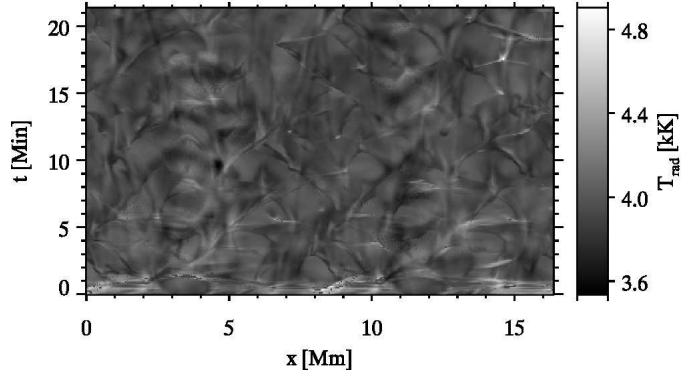}
\caption{Vertically emergent \Halpha\ line core intensity from a 2D
  radiation-MHD simulation with non-equilibrium hydrogen ionization
  taken into account. Startup effects are visible the first 3 minutes.
  There is a weak imprint of upward propagating shocks, superimposed
  on a slower-evolving pattern caused by granulation and reversed
  granulation. No parabolic outlines of DFs are present as in the
  observations \citep[see Fig. 7 of][]{2007ApJ...655..624D}.}
\label{leenaarts_fig:halpha2d}
\end{figure*}

 \begin{figure*}
\centering
\includegraphics[width=12cm]{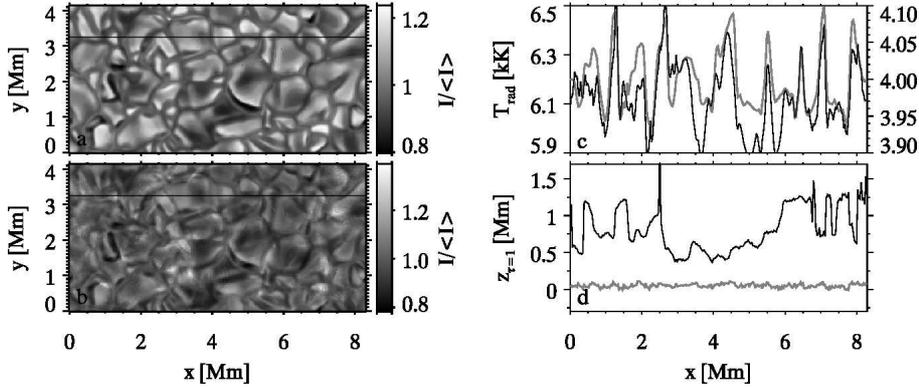}
\caption{Results of a NLTE radiative transfer computation of the
  \Halpha\ line from a 3D radiation-MHD simulation with an LTE equation
  of state. The black line in panels a
  and b indicates the cut shown in panel c and d. Panel a: vertically
  emergent intensity in the far blue wing; b: vertically emergent
  intensity at line center; c: vertically emergent intensity along the
  cut indicated in panels a and b for the blue wing (grey, left-hand
  scale) and line center (black, right hand scale); d: $\tau = 1$
  height for the blue wing (grey) and line center (black).
  \label{leenaarts_fig:halpha3d}}
\end{figure*}

The identification of DFs as magneto-acoustic shocks guided along
magnetic field lines has been established based on the excellent
agreement of dynamical properties observed in \Halpha\ with 2D
simulations
\citep{2006ApJ...647L..73H,2007ApJ...655..624D}.

In a simulation with identical resolution and magnetic field
configuration, 
\citet{2007A&A...473..625L}
investigated the effect of non-equilibrium hydrogen ionization on the
chromosphere. They found that the hydrogen $n=2$ level populations and
column densities are much higher in the simulated DFs
than elsewhere in the chromosphere. This is consistent with the
observations where DFs appear dark on top of a brighter, deeper formed
background. However, a subsequent NLTE column-by-column radiative
transfer computation employing non-equilibrium temperatures and
electron densities, but assuming statistical equilibrium for the
hydrogen level-populations, showed that the upper chromosphere is
transparent in the \Halpha\ line core in this simulation.

Fig.~\ref{leenaarts_fig:halpha2d} shows the time-evolution of the vertically
emergent \Halpha\ line center intensity. There are magnetic field
concentrations at $x=4$~Mm and $x=12$~Mm along which DFs run up in
the chromosphere. Their tops should appear as parabolic curves in this
$xt$-slice, but instead the slice shows a slowly evolving pattern of
granulation and reverse granulation. Superimposed on this pattern is a
weak imprint of upward propagating shocks (the diagonal curves
emanating from the magnetic field concentrations, for example the one
starting at $(x,t)=(5.5,10)$ Analysis of the contribution functions
shows that this imprint forms below 1~Mm, much lower than the typical
height of DFs
 \citep[see Fig. 16 of][]{2007ApJ...655..624D}.

Fig.~\ref{leenaarts_fig:halpha3d} shows the results of a NLTE column-by-column
radiative transfer computation of a snapshot of a 3D simulation of
flux-emergence by
\citet{2008ApJ...679..871M}.
This simulation used an LTE equation of state unlike the 2D simulation
that used a more realistic non-equilibrium equation of state. Panels a
and b show that the \Halpha\ line core shows granulation, even though
its $\tau = 1$ height lies between 0.3 and 1.7 Mm (panel d). The line
source function is strongly scattering and set in the upper
photosphere, where it is sensitive to the temperature pattern of the
granulation.

Both the 2D non-equilibrium simulation and the 3D LTE simulation show
an \Halpha\ line center intensity pattern that is set in the upper
photosphere, in stark contrast to the observations which show fibrilar
structure in all but the quietest areas of the atmosphere. Even in
such quiet areas the observations do not show granulation, but
signatures of shockwaves instead
\citep{2008SoPh..tmp...28R}.

Some of the reasons that might cause the absence of fibrils in the
simulated \Caline\ line core also apply here: too small computational
domain, too low spatial resolution or the limitations of the MHD
assumption. The 2D simulation necessarily had a rather simple magnetic
field topology, which might inhibit the formation of fibrils and the
mass loading of the upper chromosphere. The 3D simulation did not
suffer from this, but it had lower resolution and did not model
non-equilibrium ionization. High resolution 3D models with
non-equilibrium hydrogen ionization might have a chromosphere that is
optically thick in \Halpha.

Hydrogen line formation is however notoriously difficult to model. The
non-equilibrium ionization of hydrogen does not only affect the
electron density and temperature in the chromosphere, it also has a
strong effect on hydrogen line formation. The standard NLTE line
formation assumption of statistical equilibrium is not valid because
of the slow collisional and radiative transition rates as compared to
the hydrodynamical timescale. In addition at least the Lyman-$\alpha$
nd $\beta$ lines need to be modeled with partial frequency
redistribution (PRD) because of their strong influence on the \Halpha\
line. Both time-dependent radiative transfer and PRD effects were
ignored in the results of Figs.~\ref{leenaarts_fig:halpha2d}
and~\ref{leenaarts_fig:halpha3d}. Inclusion of these effects might significantly
increase the \Halpha\ opacity.

Thus, proper 3D modeling of the \Halpha\ line requires full
time-dependent radiative transfer with PRD in tandem with the MHD
evolution, a Herculean task that has not even been done in 1D
hydrodynamic simulations.  So far all 1D simulations perform radiative
transfer in complete redistribution
\citep{2002ApJ...572..626C,2003ApJ...589..988R,2009A&A...499..923K}.

An easier way out is to perform the 3D radiation-MHD simulations with
approximate non-equilibrium hydrogen ionization following the method
developed by 
\citet{sollum1999}
as was done in 2D by 
\citet{2007A&A...473..625L}
in order to get reasonably accurate values for the electron density
and temperature. One can then perform time-dependent radiative
radiative transfer with PRD on a series of snapshots from the
simulation to get the \Halpha\ line profiles. This separates the
detailed radiative transfer from the MHD computation, which
significantly reduces the difficulty of the problem. However, even
with this simplification it remains a daunting task to accurately
model \Halpha\ line formation in the dynamic chromosphere.

\section{Summary \& conclusions}

Multidimensional radiation-MHD simulations of the solar chromosphere
are becoming increasingly realistic. They have, amongst other things,
been used to study dynamic fibrils, flux emergence and the propagation
of Alfv\'enic waves into the corona. 

The models yet fail to reproduce some observational facts:
\begin{itemize}
\item simulated \Caline\ core images do not show fibrils;
\item the average simulated \Caline\ core is deeper and narrower than
  observed and its bisector not show the observed inverse C-shape;
\item the simulated \Halpha\ line core shows granulation, both in 2D
  simulations with non-equilibrium effects taken into accounts and in
  3D with an LTE equation of state.
\end{itemize}

In order to reproduce these the models probably require higher
resolution, larger computational domains, an improved treatment of
radiation and non-equilibrium hydrogen ionization and inclusion of
plasma effects. In order to model hydrogen transitions, in particular
\Halpha, 3D NLTE time-dependent radiative transfer codes including PRD
need to be developed. 
 
\begin{acknowledgements}
  J.L. acknowledges financial support by the European Commission through
  the SOLAIRE Network (MTRN-CT-2006-035484) and the Netherlands
  Organisation for Scientific Research (NWO).
\end{acknowledgements}

\bibliographystyle{aa}


\end{document}